\documentclass[twocolumn,pra,showpacs,superscriptaddress]{revtex4-1}
\usepackage{amsfonts}
\usepackage{amssymb}
\usepackage{amsmath}
\usepackage{graphicx}
\usepackage{epsfig}

\begin{document}

\title{$\mathcal{PT}$-symmetric trimer systems}
\author{L. Jin}
\email{jinliang@nankai.edu.cn}
\affiliation{School of Physics, Nankai University, Tianjin 300071, China}

\begin{abstract}
We studied parity-time ($\mathcal{PT}$) symmetric trimer systems that
feature open and closed boundaries. The exceptional point is three-state
coalescence at zero energy because of chiral symmetry in the open trimer;
however, two-state coalescence appears in the closed trimer affected by the
magnetic flux enclosed, which suppresses the $\mathcal{PT}$ transition.
Dynamics at the phase transition point were also qualitatively studied. For
three-state coalescence, the probability of an initial state increases in a
power law. The highest order is four; however, it can be reduced to two for
the state that is only relevant to one associated state. For two-state
coalescence in the trimer ring, we found four typical dynamical behaviors of
state probability: (i) unchanged, (ii) oscillation, (iii) quadratic
increase, and (iv) quadratic increase with oscillation. The scattering
properties of the trimer systems were also studied, particularly, asymmetric
wave emissions at spectral singularities.
\end{abstract}

\maketitle


\section{Introduction}

\label{introduction}

Parity-time ($\mathcal{PT}$) symmetric systems possess intriguing features
that derive from non-Hermiticity and the $\mathcal{PT}$ symmetry. The $%
\mathcal{PT}$-symmetric systems have been extensively investigated, both
theoretically~\cite%
{Bender98,Tateo01,Bender02,Ali02,Jones05,DNC07,DNC08,DNC2008,Znojil08,NM2008,TK2009,LJ2009,Longhi10,YNJ2010,JG2010,Brody,Philipp,HSch,PRX,Phang}
and experimentally~\cite%
{Observe,CERuter,PengNP,NatMaterFeng,CPAexp,Science2011}. The $\mathcal{PT}$%
-symmetry breaking~\cite{Observe,CERuter,PengNP}, nonreciprocal
reflectionless~\cite{NatMaterFeng}, and coherent perfect absorber~\cite%
{CPA,CPAexp,Science2011} have been observed. In 2014, $\mathcal{PT}$%
-symmetric systems have been realized by on-chip devices, known as two
coupled whispering gallery mode ring microresonators~\cite%
{Chang,PengScience,HJing}. These resonators have high quality factors, and a
gain in one resonator, induced by pumping the doped ions, can balance losses
in both resonators. The realized $\mathcal{PT}$-symmetric coupled waveguides
and resonators are mainly described by the two-site models, which act as a $%
\mathcal{PT}$-symmetric dimer~\cite{RotterRPP}. The two-site dimers are the
most concise and simple $\mathcal{PT}$-symmetric systems, however, all the $%
\mathcal{PT}$ symmetry breaking~\cite{Observe}, power oscillation~\cite%
{CERuter}, and gain induced large nonlinearity~\cite{PengNP,Chang,HJing}
were found in\ these systems. Previous study demonstrated the differences of
$\mathcal{PT}$-symmetry and pseudo-Hermiticity in non-Hermitian trimer chain
systems~\cite{Trimer}. In an optomechanical system coupled with an active
resonator, high-order EPs are useful for low-power mechanical cooling~\cite%
{HJingHighOrder}.\ Triple-cavity supermodes was analysised based on thye
coupled mode theory~\cite{MXiao2017}. Besides, $\mathcal{PT}$-symmetric
oligomers were investigated in nonlinearity systems, the solution and
nonlinear dynamics were examined~\cite{PRE2011,PTRSA2013}. The asymmetric
scattering properties for left and right propagation were demonstrated for
nonlinear $\mathcal{PT}$-symmetric oligomers embedded in linear lead; the
plane waves are dynamically unstable except in the vicinity of the linear
limit; however, the asymmetric transmissions persist for Gaussian wave
packets~\cite{JPA2012}.

This paper focuses\ on the $\mathcal{PT}$-symmetric trimer. We investigated
two $\mathcal{PT}$-symmetric trimer systems under different boundary
conditions (one with an open chain form and the other with a closed ring
form) and analytically solved their Hamiltonians. We concentrate on dynamics
at the $\mathcal{PT}$-symmetric phase transition (i.e., exceptional) point
of trimer systems, particularly in the presence of effective magnetic flux.
From the eigen spectrum, we obtained the $\mathcal{PT}$-symmetric phase
transition point and revealed that the uniformly coupled trimer chain with
balanced gain and loss at its ends has one exceptional point, where three
eigenstates coalesce at zero energy. In the trimer ring, a nonreciprocal
coupling connects the head and tail of the trimer chain and the
nonreciprocal coupling factors correspond to a gauge invariant field, which
can be seen as an effective magnetic flux. For the trimer ring, the $%
\mathcal{PT}$-symmetric phase transition point can be a two or three
eigenstates coalescence. To determined the dynamical features at the
exceptional point in the trimer systems, we examined the time evolution of
different initial states. For the trimer chain, the probability increases at
the exceptional point in a power law. Because of the three-state coalescence
(EP3)~\cite{EP3}, the highest order in the power law is four; however, this
probability increases quadratically at the EP3 when the initial state is
only related to the eigenstate and the first associated state. In the trimer
ring, the $\mathcal{PT}$-symmetric phase transition point is a two-state
coalescence (EP2) except for the effective magnetic flux, which is $\Phi
=2n\pi +\pi /2$ ($n\in \mathbb{Z}$). For the EP2, the coalesced two
eigenenergy and the third one are all nonzero. The evolved probability of an
initial state has four typical behaviors, (i) unchanged, (ii) oscillation,
(iii) quadratic increase, and (iv) quadratic increase with oscillation.
Notably, the oscillation periods are the same, because they are coalesced
eigenenergy dependent, i.e., $2\pi /E_{2}$. We also investigated the
scattering properties of the trimer systems, particularly, the spectral
singularities and asymmetric wave emissions under the influence of
nonreciprocal coupling.

The remainder of this paper is organized as follows. In Sec.~\ref{Model}, we
present the trimer systems, the $\mathcal{PT}$-symmetric phase diagram, and
the spectrum. In Sec.~\ref{EP}, we investigate the dynamics at the
exceptional point. In Sec.~\ref{SS}, we explore the scattering properties of
the trimer systems, especially the spectral singularities and asymmetric
wave emission. Finally in Sec.~\ref{Summary}, we summarize the results and
conclude our study.

\section{$\mathcal{PT}$-symmetric trimer}

\label{Model}

\begin{figure}[tbp]
\includegraphics[bb=0 0 450 150, width=8.4 cm, clip]{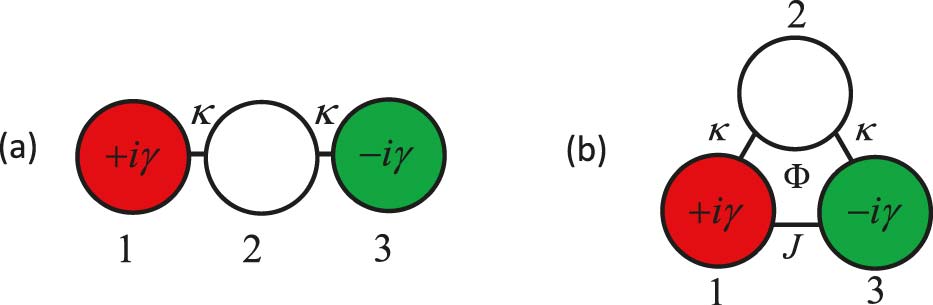}
\caption{(Color online) Schematic configuration of trimer systems. (a) A
trimer chain with open boundary. (b) A trimer ring with periodical boundary.
The trimer systems have a pair of balanced gain (red) and loss (blue).} %
\label{fig1}
\end{figure}

\textit{Open boundary.}---The open boundary trimer system has three sites
coupled in chain form, with a coupling strength of $\kappa $. Specially, the
trimer has a gain and loss site coupled to an intermediary site between
them. The system is schematically illustrated in Fig.~\ref{fig1}(a), and the
trimer chain Hamiltonian is
\begin{equation}
H_{0}=-\kappa (a_{1}^{\dagger }a_{2}+a_{2}^{\dagger }a_{3}+\mathrm{h.c.}%
)+i\gamma a_{1}^{\dagger }a_{1}-i\gamma a_{3}^{\dagger }a_{3},
\end{equation}%
where $a_{j}^{\dagger }$ ($a_{j}$) is the creation (annihilation) operator
of site $j$ and $\mathcal{P}$ is the parity operator, which satisfies $%
\mathcal{P}a_{j}^{\dagger }\mathcal{P}^{-1}=a_{4-j}^{\dagger }$, $\mathcal{P}%
a_{j}\mathcal{P}^{-1}=a_{4-j}$; $\mathcal{T}$ is the time-reversal operator,
which satisfies $\mathcal{T}i\mathcal{T}^{-1}=-i$. According to the
definitions of the $\mathcal{P}$ and $\mathcal{T}$ operators, the trimer
chain Hamiltonian is $\mathcal{PT}$-symmetric, i.e., $\left( \mathcal{PT}%
\right) H_{0}\left( \mathcal{PT}\right) ^{-1}$. We can also define a unitary
transformation $S$ through $Sa_{j}^{\dagger }S^{-1}=(-1)^{j}a_{4-j}^{\dagger
}$ and $Sa_{j}S^{-1}=(-1)^{j}a_{4-j}$, where the trimer chain Hamiltonian
under the unitary transformation satisfies $SH_{0}S^{-1}=-H_{0}$. This
indicates that the trimer chain also has a chiral symmetry, and that the
spectrum is symmetrical at zero energy. The eigenvalues found are $0$ and $%
\pm \sqrt{2\kappa ^{2}-\gamma ^{2}}$. All three eigenstates are $\mathcal{PT}
$-symmetric for balanced gain and loss in the region $|\gamma /\kappa |<%
\sqrt{2}$, and the trimer chain is in the exact $\mathcal{PT}$-symmetric
phase. When $|\gamma /\kappa |>\sqrt{2}$, the $\mathcal{PT}$ symmetry of two
eigen states with imaginary energy $\pm i\sqrt{\gamma ^{2}-2\kappa ^{2}}$
breaks, and the trimer chain enters the broken $\mathcal{PT}$-symmetric
phase. By contrast, when $|\gamma /\kappa |=\sqrt{2}$, the trimer chain is
at the exceptional point, which is the $\mathcal{PT}$-symmetric phase
transition point~\cite{HeissEP12}, where three eigenstates coalesce with
zero energy. Notably, this is the only exceptional point in the trimer chain.

\textit{Periodical boundary.}---A three-site system can also constitute a
trimer coupled together in ring form, as schematically illustrated in Fig.~%
\ref{fig1}(b). The Hamiltonian is denoted as $H$ in the following:
\begin{equation}
H=(-\kappa a_{1}^{\dagger }a_{2}-\kappa a_{2}^{\dagger }a_{3}-Je^{i\Phi
}a_{1}^{\dagger }a_{3}+\mathrm{h.c.})+i\gamma a_{1}^{\dagger }a_{1}-i\gamma
a_{3}^{\dagger }a_{3},
\end{equation}%
which has an additional nonreciprocal hopping $-Je^{\pm i\Phi }$ between
sites $1$ and $3$ compared with the trimer chain. Moreover, $H$ can be
expressed as $H=H_{0}-Je^{i\Phi }a_{1}^{\dagger }a_{3}-Je^{-i\Phi
}a_{3}^{\dagger }a_{1}$. The nonreciprocal phase factor $e^{\pm i\Phi }$ is
equivalent to an effective magnetic flux $\Phi $ enclosed in the trimer
ring, which can be introduced in the coupling process through an optical
path imbalance method~\cite{Hafezi}. The trimer ring retains identical $%
\mathcal{PT}$ symmetry in the presence of the nonreciprocal hopping term;
however, the trimer ring is chirally symmetric at magnetic flux $\Phi =n\pi
+\pi /2$ ($n\in \mathbb{Z}$), i.e., $SHS^{-1}=-H$. The nonreciprocal
coupling enriches the varieties of the trimer spectrum. In particular, the $%
\mathcal{PT}$-symmetric transition of the trimer changes considerably.

The trimer ring is a $3\times 3$ matrix. Under the basis $\{a_{1}^{\dagger
}\left\vert \mathrm{vac}\right\rangle ,a_{2}^{\dagger }\left\vert \mathrm{vac%
}\right\rangle ,a_{3}^{\dagger }\left\vert \mathrm{vac}\right\rangle \}$, $H$
is in the following form:
\begin{equation}
\left(
\begin{array}{ccc}
i\gamma & -\kappa & -Je^{i\Phi } \\
-\kappa & 0 & -\kappa \\
-Je^{-i\Phi } & -\kappa & -i\gamma%
\end{array}%
\right) .
\end{equation}%
The eigenenergy can be analytically determined from the cubic equation $%
E^{3}+\left( \gamma ^{2}-J^{2}-2\kappa ^{2}\right) E+2J\kappa ^{2}\cos \Phi
=0$. The $\mathcal{PT}$ symmetry transition point occurs at
\begin{equation}
\left( \gamma ^{2}-J^{2}-2\kappa ^{2}\right) ^{3}/3^{3}+\left( 2J\kappa
^{2}\cos \Phi \right) ^{2}/2^{2}=0,
\end{equation}%
which indicates the exceptional points of the trimer ring. Unlike the trimer
chain, the exceptional points in the trimer ring can be a coalescence of two
eigenstates for effective magnetic flux that is not at half magnetic flux
quantum ($\Phi \neq n\pi +\pi /2$, $n\in \mathbb{Z}$).

\begin{figure}[tbp]
\includegraphics[bb=10 0 510 490, width=8.4 cm, clip]{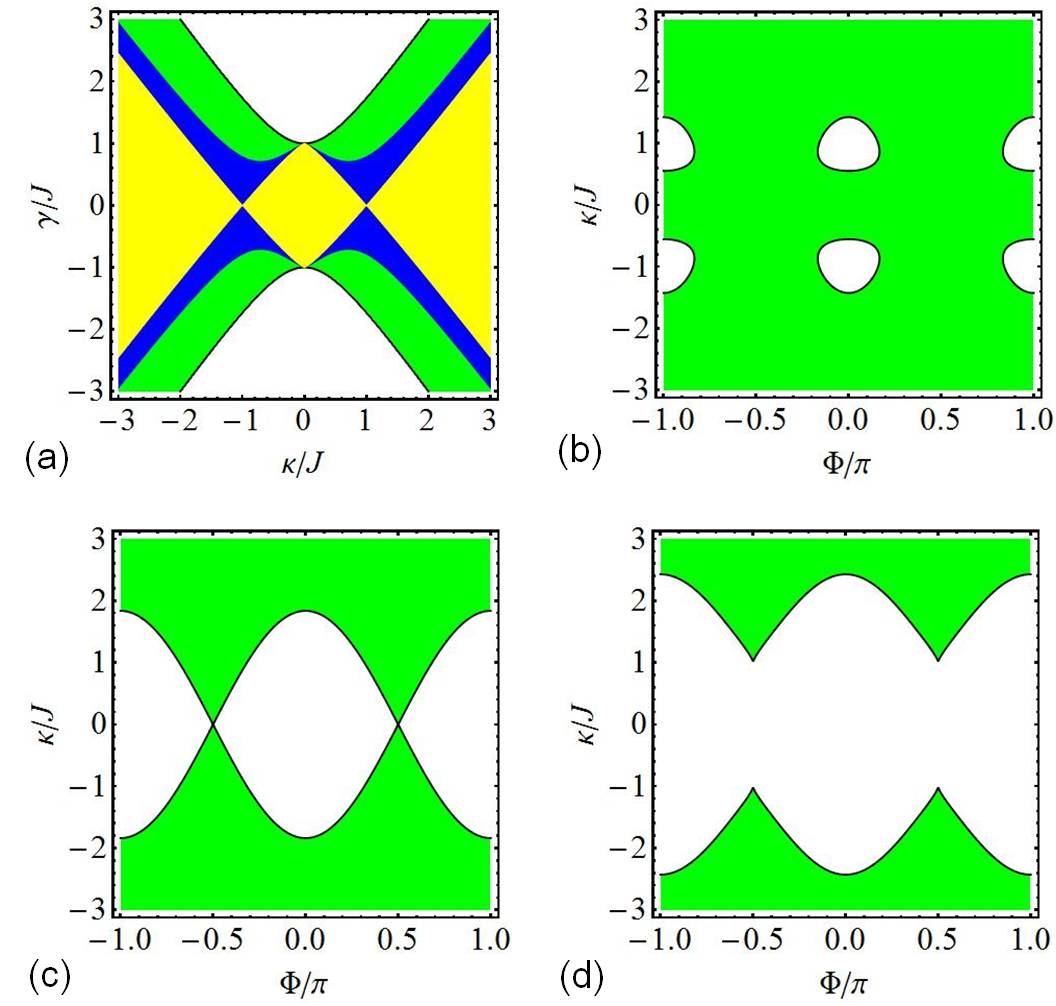}
\caption{(Color online) Magnetic flux suppresses the $\mathcal{PT}$ transition. (a) Phase diagram of the trimer ring in the $\protect\kappa$ and $\protect\gamma$ plane. The yellow region is the $\mathcal{PT}$-symmetric phase for $\Phi=0$. The $\mathcal{PT}$-symmetric region becomes
large as magnetic flux increases, the blue region is the additional $\mathcal{PT}$-symmetric region for $\Phi=\protect\pi/4$. All the colored
regions are for $\Phi=\protect\pi/2$, which shows the $\mathcal{PT}$-symmetric region at its maximum. (b-d) Phase diagram of the trimer ring in
the $\Phi$ and $\protect\kappa$ plane. The green regions are the $\mathcal{PT}$-symmetric phase for (b) $\protect\gamma/J=1/2$, (c) $\protect\gamma/J=1$, and
(d) $\protect\gamma/J=\protect\sqrt{3}$.} \label{fig2}
\end{figure}

In Fig.~\ref{fig2}, the phase diagram is depicted as a function of coupling
strength and gain rate. In the plots, the regions filled with colors
represent the exact $\mathcal{PT}$-symmetric phase for different magnetic
fluxes. At trivial magnetic flux $\Phi =2n\pi $ ($n\in
\mathbb{Z}
$), the $\mathcal{PT}$-symmetric phase is portrayed in the yellow areas. At
coupling $\kappa =0$, the trimer ring reduces into a dimer chain with a sole
site, and the critical $\mathcal{PT}$ symmetry transition point for the
dimer is well known at $\gamma _{\mathrm{c}}=J$. However, at coupling $%
\kappa =J$, the trimer is a uniform coupled ring, and the $\mathcal{PT}$
symmetry is fragile; any nonzero gain or loss ($\gamma \neq 0$) breaks the $%
\mathcal{PT}$ symmetry, and eigenenergies of a pair of eigenstates become
complex conjugation. \textit{Magnetic flux suppresses the} $\mathcal{PT}$%
\textit{\ transition}. In the presence of nontrivial magnetic flux $\Phi
\neq 2n\pi $ ($n\in
\mathbb{Z}
$), the $\mathcal{PT}$-symmetric region enlarges, and the blue areas
indicate the additional $\mathcal{PT}$-symmetric region at $\Phi =2n\pi +\pi
/4$ compared with $\Phi =2n\pi $ ($n\in
\mathbb{Z}
$). In this situation, at coupling $\kappa =0$, the trimer ring reduces into
a dimer chain and a sole site, and the critical point is also at $\gamma _{%
\mathrm{c}}=J$. This reflects how nonreciprocal phase factor $e^{\pm i\Phi }$
has no effect in the dimer chain. By taking a local transformation $%
a_{1}^{\dagger }\rightarrow e^{i\Phi }a_{1}^{\dagger }$, $a_{1}\rightarrow
e^{-i\Phi }a_{1}$ or $a_{3}\rightarrow e^{i\Phi }a_{3}$, $a_{3}^{\dagger
}\rightarrow e^{-i\Phi }a_{3}^{\dagger }$, the nonreciprocal phase factor $%
e^{\pm i\Phi }$ in the couplings $Je^{\pm i\Phi }$ is removed. When magnetic
flux is $\Phi =2n\pi +\pi /2$ ($n\in
\mathbb{Z}
$), the $\mathcal{PT}$-symmetric region has additional areas highlighted in
green compared with $\Phi =2n\pi +\pi /4$ ($n\in
\mathbb{Z}
$). The trimer ring has the largest region with exact $\mathcal{PT}$%
-symmetric phase. All of the areas in Fig.~\ref{fig2} filled with colors
represent the exact $\mathcal{PT}$-symmetric phase, whereas the white areas
are the broken $\mathcal{PT}$-symmetric phase for half magnetic flux quantum
$\Phi =2n\pi +\pi /2$ ($n\in \mathbb{Z}$).

The critical balanced gain and loss rate is the smallest at coupling $\kappa
=0$ and it increases as the coupling $\kappa $ in the form of $\gamma _{%
\mathrm{c}}=\sqrt{J^{2}+2\kappa ^{2}}$. At this $\mathcal{PT}$-symmetric
phase transition point ($\gamma _{\mathrm{c}}=\sqrt{J^{2}+2\kappa ^{2}}$),
three eigenstates of the trimer ring coalesce (except for the trimer reduced
to a dimer at coupling $\kappa =0$ ($\gamma _{\mathrm{c}}=J$), which is an
EP2). In Figs.~\ref{fig2}(b-d), the $\mathcal{PT}$-symmetric phase diagram
in the parameter plane of $\Phi $ and $\kappa $ is plotted. For $|\gamma
/J|<0$, as shown in Figs.~\ref{fig2}(b,c), we acquire four typical rules:
(i) the $\mathcal{PT}$-symmetric phase is the most fragile at magnetic flux $%
\Phi =n\pi $ ($n\in
\mathbb{Z}
$) and $\kappa /J=1$; (ii) the exact $\mathcal{PT}$-symmetric region widens
as the magnetic flux increases; (iii) at coupling $\kappa =0$, magnetic flux
does not affect the $\mathcal{PT}$-symmetric of the trimer ring, and the
trimer ring is in the exact $\mathcal{PT}$-symmetric phase; and (iv) the $%
\mathcal{PT}$ symmetry is robust at magnetic flux $\Phi =n\pi +\pi /2$ ($%
n\in
\mathbb{Z}
$) and the trimer ring is in the exact $\mathcal{PT}$-symmetric phase. All
four rules remain unchanged as $\gamma $ increases from $0$ to $J$; however,
the exact $\mathcal{PT}$-symmetric region shrinks when $\gamma $ increases.
For weak gain and loss $\gamma <J$, the trajectory of exceptional point
forms an island with broken $\mathcal{PT}$ symmetry [Fig.~\ref{fig2}(b)].
The isolated broken $\mathcal{PT}$-symmetric region in parameter space was
previously found in a dimerized photonic crystals~\cite{CTChanPRB}, where
two kinds of $\mathcal{PT}$ transitions were found: (i) reentering exact $%
\mathcal{PT}$-symmetric phase from broken $\mathcal{PT}$-symmetric phase at
higher non-Hermiticity; and (ii) coalescence of EPs from the Brillouin zone
center and boundary generates higher order of EPs in the interior of the
Brillouin zone~\cite{CTChanPRB}. By contrast, we have similar conclusions
related to parameters $\kappa $ and $\Phi $. There are also two kinds of $%
\mathcal{PT}$ transitions in the trimer ring. (i) As the coupling strength $%
\kappa $ increases, the system crosses broken $\mathcal{PT}$-symmetric phase
from exact $\mathcal{PT}$-symmetric phase and reenters it [Fig.~\ref{fig2}%
(b)]; (ii) EPs from magnetic flux $\Phi =0$ and $\Phi =\pi $ coalesce and
form higher order EPs at $\Phi =\pm \pi /2$ [Figs.~\ref{fig2}(c) and~\ref%
{fig2}(d)]. At gain and loss $\gamma =J$, the trimer ring is an EP2
exceptional point at $\kappa =0$ for any magnetic flux, and the trimer ring
is in the exact $\mathcal{PT}$-symmetric phase at $\Phi =n\pi +\pi /2$ for $%
\kappa \neq 0$ [Fig.~\ref{fig2}(c)]. However, for $|\gamma /J|>1$, the $%
\mathcal{PT}$-symmetric is most robust at $\Phi =n\pi +\pi /2$ ($n\in
\mathbb{Z}
$). The trimer is $\mathcal{PT}$-symmetric at $|\kappa /J|>1$ for $\gamma/J=%
\sqrt{3}$. As Fig.~\ref{fig2}(d) reveals, the trimer ring is in the broken $%
\mathcal{PT}$-symmetric region for coupling $|\kappa /J|<1$ and in the exact
$\mathcal{PT}$-symmetric region for $|\kappa /J|>2.43$. Additionally, the $%
\mathcal{PT}$-symmetric region expands as magnetic flux increases from $\Phi
=2n\pi $ to $\Phi =2n\pi +\pi /2$ ($n\in
\mathbb{Z}
$) for the coupling in between $1<|\kappa /J|<2.43$, and subsequently
shrinks as magnetic flux increases from $\Phi =2n\pi +\pi /2$ to $\Phi
=(2n+1)\pi $ ($n\in
\mathbb{Z}
$).

\begin{figure}[tb]
\includegraphics[bb=0 0 470 200, width=8.4 cm, clip]{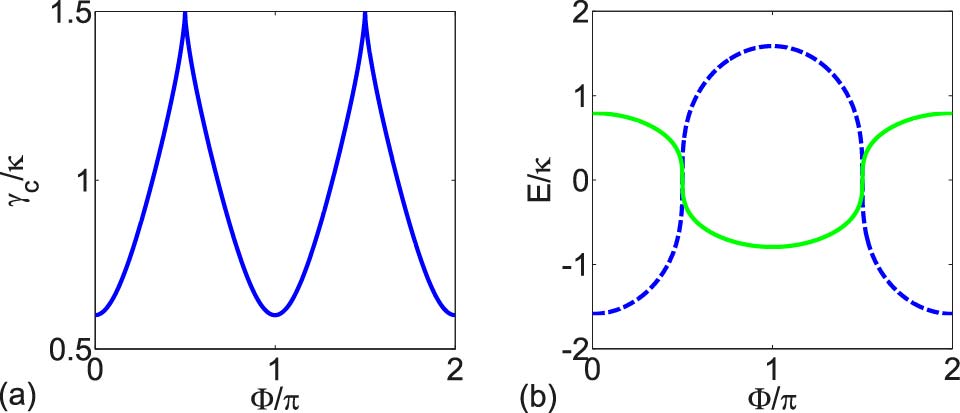}
\caption{(Color online) (a) Critical gain and loss of Eq.~\protect\ref{EP2}
at EP2, and (b) the spectrum at EP2 as functions of magnetic flux $\Phi$.
The green dotted line in (b) represents the two coalesced energies, and the
parameters are $\protect\kappa=1$ and $J=1/2$.} \label{fig3}
\end{figure}

In the $\mathcal{PT}$-symmetric system, the coalesced states encountered are
mostly the extensively investigated EP2, which leads to a probability
increase under a power law with a highest order of two, $\sum_{n=0}^{2}%
\alpha _{n}t^{n}$~\cite{MoiseyevEP11}. In the open chain form trimer shown
in Fig.~\ref{fig1}(a), the only exceptional point is an EP3 at $\gamma _{%
\mathrm{c}}=\sqrt{2}\kappa $. EP3 is a high-order coalescence, where the
coalesced states consist of one eigenstate and two associated states. The
probability in the system increases under a power law, and the evolved
probability is a polynomial function of time, i.e., $\sum_{n=0}^{4}\alpha
_{n}t^{n}$. Notably, the highest order in the polynomial for EP3 is four.
However, in the close ring form trimer shown in Fig.~\ref{fig1}(b), two
types of coalescence may occur: an EP3 when $\Phi =2n\pi +\pi /2$ ($n\in
\mathbb{Z}
$) and $\gamma _{\mathrm{c}}=\sqrt{J^{2}+2\kappa ^{2}}$, or an EP2 when $%
\Phi \neq 2n\pi +\pi /2$ ($n\in
\mathbb{Z}
$). Critical gain and loss for the EP2 is
\begin{equation}
\gamma _{\mathrm{c}}=\sqrt{2\kappa ^{2}+J^{2}-3\sqrt[3]{\left( J\kappa
^{2}\cos \Phi \right) ^{2}}},  \label{EP2}
\end{equation}%
which is plotted in Fig.~\ref{fig3}(a) for coupling strengths $\kappa =1$
and $J=1/2$ as a function of the effective magnetic flux $\Phi $. The energy
spectrum at EP2 is plotted in Fig.~\ref{fig3}(b), where two states of three
are coalesced; however, at magnetic flux $\Phi =\pi /2$, all three states
coalesced. The spectrum is central symmetric at zero energy. When magnetic
flux varies from $\Phi _{0}$ to $\pi -\Phi _{0}$, the spectrum $E_{1,2}$ is
inversed to $-E_{1,2}$. As depicted in Fig.~\ref{fig2}(a), the boundaries
between regions with different colors indicate when EP2 occurs, whereas the
boundaries (black curves) between the colored regions and the white region
indicate when EP3 occurs.

\begin{figure}[tb]
\includegraphics[bb=10 0 445 610, width=8.4 cm, clip]{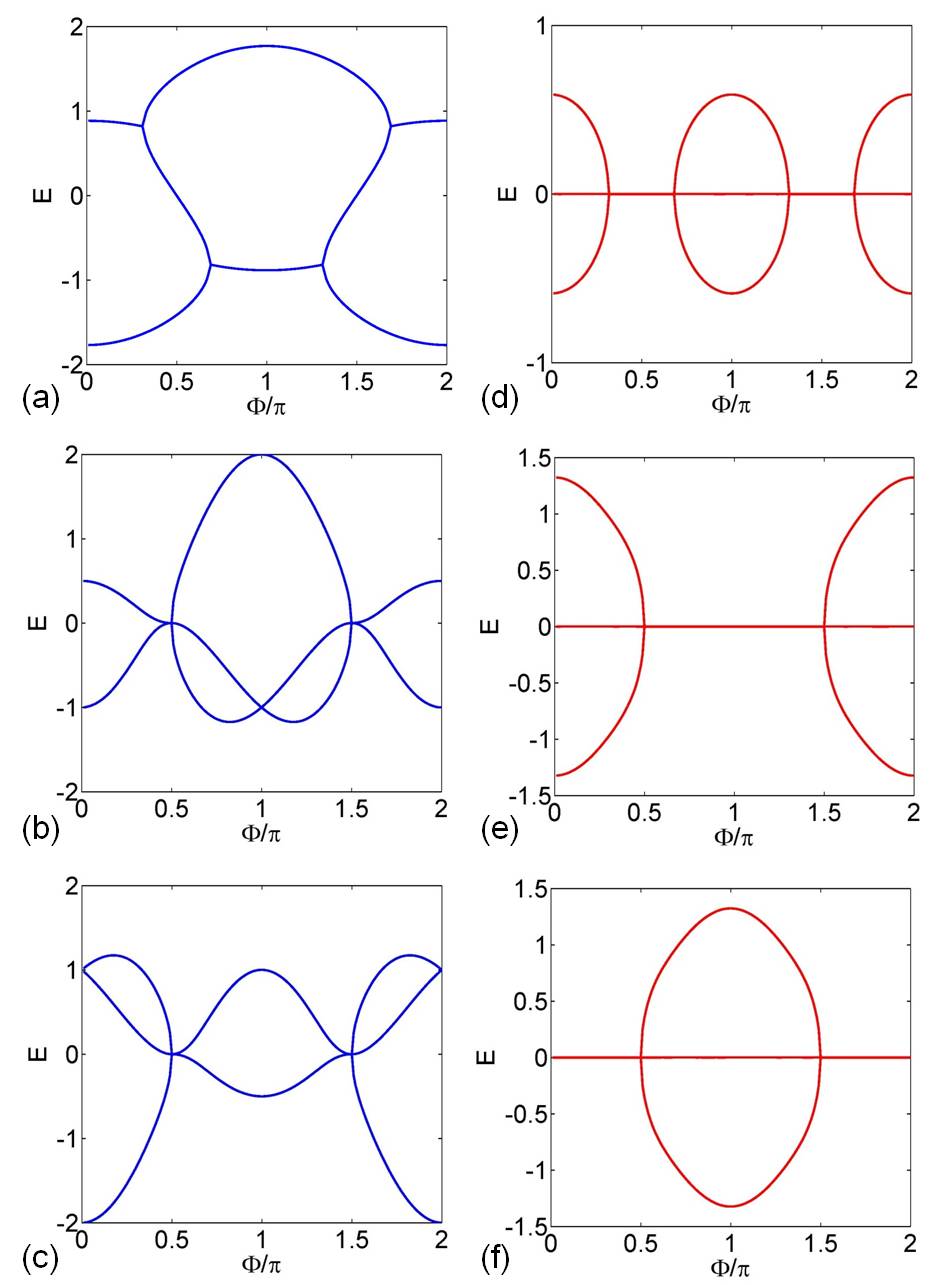}
\caption{(Color online) (a-c) Real part (in blue) of the trimer ring
spectrum. (d-f) Imaginary part (in red) of the trimer ring spectrum. The
parameters are (a,d) $J=\protect\kappa=1$, (b,e) $\protect\gamma /J=1+\cos
\Phi $, $\protect\kappa /J=\cos \Phi $, and (c,f) $\protect\gamma /J=1-\cos
\Phi $, $\protect\kappa /J=\cos \Phi $.} \label{fig4}
\end{figure}

The eigenenergies are solved from the cubic equation, and the expressions
are complicated. We depict the real and imaginary part of the three band in
Fig.~\ref{fig4} for several cases. Specifically, the bands are symmetric
with respect to $\Phi =\pi $. In Figs.~\ref{fig4}(a) and~\ref{fig4}(d), $%
\gamma =J=\kappa =1 $ [along $\kappa /J=1$ in Fig.~\ref{fig2}(c)], the $%
\mathcal{PT}$-symmetric regions in $\Phi \in \left[ 0,2\pi \right] $ are $%
1<\Phi <\pi -1$ and $\pi +1<\Phi <2\pi -1$. As effective magnetic flux
changes from $\Phi =0$ to $\Phi =1 $, the two energies with higher real part
are the complex conjugate pairs; they change to real energy at $\Phi =1$ and
the system is in the exact $\mathcal{PT}$-symmetric region during the shift
from $\Phi=1$ to $\Phi=\pi-1$ with three real energies. The states with the
highest and lowest energies continue increasing in this region until $\Phi
=\pi -1$, where the state with the mediate real energy coalesces with the
state with the lowest real energy. Subsequently, from $\Phi =\pi -1$ to $%
\Phi =\pi $, the higher real energy increases, but the lower two real
energies decrease. The conjugate pair is the two states with lower real
energy. As $\Phi $ changes in the region $\left[ 0,\pi \right] $, the trimer
ring transitions from broken $\mathcal{PT}$-symmetric to exact $\mathcal{PT}
$-symmetric and back to broken $\mathcal{PT}$-symmetric again. The $\mathcal{%
PT}$ symmetry transition point is an EP2, and the coalesced energy is
nonzero and $\Phi $ dependent. In Figs.~\ref{fig4}(b) and~\ref{fig4}(e), the
parameters are given as $\gamma /J=1+\cos \Phi $ and $\kappa /J=\cos \Phi $;
in this situation, the trimer ring is in the exact $\mathcal{PT}$-symmetric
phase for $\pi /2<\Phi <3\pi /2$. By contrast, in Figs.~\ref{fig4}(c) and~%
\ref{fig4}(f), the parameters are given as $\gamma /J=1-\cos \Phi $ and $%
\kappa /J=\cos \Phi $, and the trimer ring is in the broken $\mathcal{PT}$%
-symmetric phase for $\pi /2<\Phi <3\pi /2$.

As all of these figures [Figs.~\ref{fig4}(b),~\ref{fig4}(c),~\ref{fig4}(e),
and~\ref{fig4}(f)]\ reveal, the trimer ring experiences three states
coalesced $\mathcal{PT}$ symmetry transition points as the parameters
change. For $\Phi =\pi $ in Figs.~\ref{fig4}(b) and~\ref{fig4}(e), the
trimer ring is a Hermitian trimer ring with antisymmetric periodical
boundary condition at the gain and loss $\gamma =0$. The energy degeneracy
is the traditional degeneracy in Hermitian systems, the same situation which
occurs in Figs.~\ref{fig4}(c) and~\ref{fig4}(f) at $\Phi =0$.

\section{Dynamics at exceptional point}

\label{EP}

The exceptional points universally exist in non-Hermitian systems~\cite%
{HeissEP04Cz,HeissEP04}. Their existence and the role they play for the
dynamics of open quantum systems, were investigated by studying the
effective Hamiltonian using the Feshbach projection technique. Specifically,
the topological structures of the exceptional points significantly affect
the dynamical properties~\cite{RotterEP08,RotterEPJPA}, and the dynamics at
the exceptional points have been investigated in many quantum systems~\cite%
{NM13,Longhi14,NM15}. Because of the unique features of the exceptional
points, they have been widely applied to sensitivity enhancement~\cite%
{Wiersig}, parameter estimation~\cite{NMNJP,NMPRA}, and topological energy
transfer~\cite{Harris16}.

\textit{Calculation of time evolution.}---The Hamiltonian is
nondiagonalizable at the exceptional point; instead, changed into a Jordan
block form after transformation, namely, $VHV^{-1}=h$, where $h$ is a Jordan
block. Generally, $h$ comprises diagonal components (formed by the
eigenstates) and Jordan block components (formed by the coalesced states);
in other words, $h$ is diagonalized, except for the nondiagonalizable Jordan
block components. The differential equations (i.e., the Schr\"{o}dinger
equations) can be solved directly by $i\mathrm{d}\psi \left( t\right) /%
\mathrm{d}t=h\psi \left( t\right) $ with $\psi \left( 0\right) =V^{-1}\Psi
\left( 0\right) $ and the initial state $\Psi \left( 0\right) $. The
diagonalized components correspond to the eigenstates of $H$, and the time
evolution is simply a superposition of the eigenstates with dynamical factor
$e^{-iEt}$, where $E$ is the eigenenergy. For the Jordan block components,
the coalesced states are linked in the differential equations, which results
in the evolution state in power law of time and the increase of probability.
The time evolution can be obtained from $\Psi \left( t\right) =V\psi \left(
t\right) $. Further details about the calculation of time evolution are
provided in Appendix.

\begin{figure}[tb]
\includegraphics[bb=0 0 450 215, width=8.2 cm, clip]{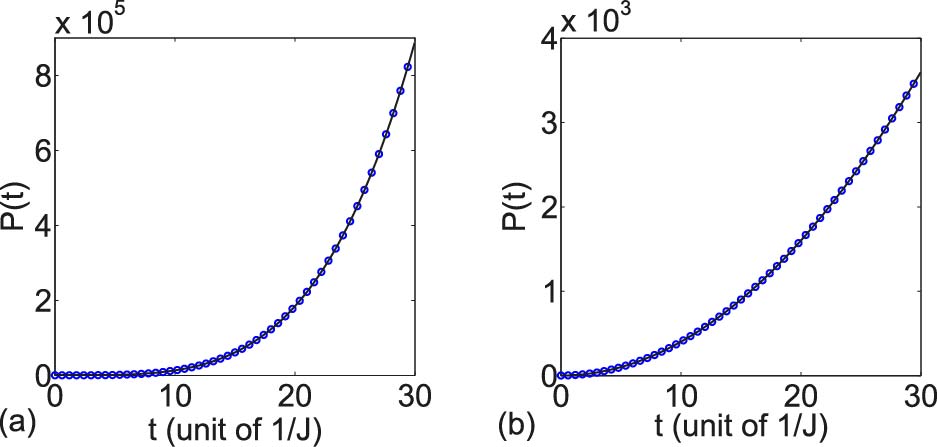}
\caption{(Color online) Time evolution probability of the trimer at EP3. (a)
Initial state is $\left\vert \Psi (0)\right\rangle =[1,0,0]^{T}$. (b)
Initial state is $\left\vert \Psi (0)\right\rangle =[1/\protect\sqrt{2},0,1/\protect\sqrt{2}]^{T}$. The trimer parameters are $\protect\kappa =1$ and $\protect\gamma =\protect\sqrt{2}$, and the black lines (blue crosses) are
the analytical (numerical) results.} \label{fig5}
\end{figure}

\textit{Open boundary trimer.}---We first considered the trimer chain. As
discussed in Sec.~\ref{Model}, the trimer has a pair of symmetric energy, $%
\pm \sqrt{2\kappa ^{2}-\gamma ^{2}}$, and a parameter independent zero
energy state protected by chiral symmetry; the trimer is at the exceptional
point when $\gamma =\sqrt{2}\kappa $. In this situation, three states
coalesce at energy zero and the only eigenstate in $\mathcal{PT}$-symmetric
form is $(1/2)[-i,\sqrt{2},i]^{T}$. The time evolution is obtained directly
by solving the differential equations from the Schr\"{o}dinger equations
(see Appendix). The coalesced states induce a polynomial power increase of
the initial state probability. For instance, the probability of an initial
state $\Psi \left( 0\right) =[1,0,0]^{T}$ is $P(t)=\left\vert \Psi \left(
t\right) \right\vert ^{2}=1+2\sqrt{2}\left( \kappa t\right) +4\left( \kappa
t\right) ^{2}+2\sqrt{2}\left( \kappa t\right) ^{3}+\left( \kappa t\right)
^{4}$ as shown in Fig.~\ref{fig5}(a). Here, the initial state is related to
the eigenstate and its two associated states, which results in the evolved
amplitude of the state growing quadratically. Thus, the probability is a
function of time in polynomial form $\sum_{n=0}^{4}\alpha _{n}\left( \kappa
t\right) ^{n}$. During a long time interval, the power law probability
increase is the highest order dominant, i.e., $P(t)\approx \alpha _{4}\left(
\kappa t\right) ^{4}$. However, in special situations, an initial state may
relevant to the eigenstate and one of its associated states, where the
evolved amplitude of the state can linearly increase as time progresses.
Therefore, the highest power order of probability increase can be reduced to
two. The condition for a quadratical probability increase at EP3 is $\Psi
_{1}\left( 0\right) +i\sqrt{2}\Psi _{2}\left( 0\right) -\Psi _{3}\left(
0\right) =0$. For instance, we consider an initial state $\Psi \left(
0\right) =[1/\sqrt{2},0,1/\sqrt{2}]^{T}$, where the evolved probability is $%
P(t)=1+4\left( \kappa t\right) ^{2}$ [Fig.~\ref{fig5}(b)].

\textit{Periodical boundary trimer.}---In the following, we discuss the
dynamics of EP3 and EP2 in detail. We considered the EP3 of the closed
trimer, which requires the effective flux at a quarter quantum flux $\Phi
=2n\pi +\pi /2$ ($n\in
\mathbb{Z}
$). In this situation, the trimer is chiral symmetric, and three eigenstates
coalesce at energy zero. The trimer Hamiltonian is nondiagonalizable but can
be reduced to a $3\times 3$ Jordan block. The only eigenenergy is zero, and
the eigenstate is $\Lambda ^{-1/2}[%
\begin{array}{ccc}
-i\kappa & J+\sqrt{J^{2}+2\kappa ^{2}} & i\kappa%
\end{array}%
]^{T}$, where $\Lambda $ is the renormalization factor $\Lambda =2\sqrt{%
J^{2}+2\kappa ^{2}}(\sqrt{J^{2}+2\kappa ^{2}}+J)$. In Appendix, we describe
the procedure by which we obtained the time evolution dynamics. Generally,
the initial state probability as a function of time is a polynomial form
with no orders larger than four. To interpret the dynamics in detail, we
considered a concrete trimer ring with couplings $\kappa =1$ and $J=1/2$,
and the critical gain and loss of $\gamma _{\mathrm{c}}=3/2$ under an
effective magnetic flux $\Phi =\pi /2$. The only eigenstate for the trimer
ring is $(1/\sqrt{6})[%
\begin{array}{ccc}
-i & 2 & i%
\end{array}%
]^{T}$. Thus, the evolved probability calculated is $P(t)=1+3\left( \kappa
t\right) +(9/2)\left( \kappa t\right) ^{2}+3\left( \kappa t\right)
^{3}+(3/2)\left( \kappa t\right) ^{4}$ initially for an excitation on the
gain site, namely, the state $\Psi \left( 0\right) =[1,0,0]^{T}$. The
probabilities are determined by the evolved state amplitude and can change
to other forms for some special initial states. The increases vary in
manners; for example, an excitation on the central site (without gain or
loss) is $\Psi \left( 0\right) =[0,1,0]^{T}$ and the probability increase is
in an exact quartic form, $P(t)=1+\chi \left( \kappa t\right) ^{4}$, where
the factor $\chi $ is coupling strength dependent
\begin{equation}
\chi =\frac{2\kappa ^{2}+J^{2}+J\gamma _{\mathrm{c}}}{\kappa
^{2}+J^{2}+J\gamma _{\mathrm{c}}}.
\end{equation}%
When $\kappa =1$ and $J=1/2$, the factor is $\chi =3/2$, and the probability
is $P(t)=1+(3/2)\left( \kappa t\right) ^{4}$.

Notably, the probability increase at EP3 can be reduced to a quadratic form,
just like system at an EP2. For the initial state $\Psi \left( 0\right) =(1/%
\sqrt{2})[1,0,1]^{T}$, we calculated the evolved probability, which is $%
P(t)=1+3\left( \kappa t\right) ^{2}$. The condition for a reduced\ order in
the power law probability increase is $\Psi _{1}\left( 0\right) +i\Psi
_{2}\left( 0\right) -\Psi _{3}\left( 0\right) =0$ for the trimer ring of $%
\kappa =1$, $J=1/2$, and $\gamma _{\mathrm{c}}=3/2$ at $\Phi =2n\pi +\pi /2$
($n\in
\mathbb{Z}
$); and the condition changes to $\Psi _{1}\left( 0\right) +2i\Psi
_{2}\left( 0\right) -\Psi _{3}\left( 0\right) =0$ at $\Phi =2n\pi -\pi /2$ ($%
n\in
\mathbb{Z}
$). In these cases, the highest power of probability increase is two, and it
increases monotonically at EP3 for the three eigenstates coalescence in the
trimer ring. For the trimer ring at EP2 with two eigenstates coalescence,
the probability generally increases and oscillates. The oscillation is
because of the influence of the third real energy eigenstate. However,
special states that are relevant only to the two coalesced states also exist
and their probabilities increase monotonically.

\begin{figure}[tb]
\includegraphics[bb=0 0  475 430, width=8.6 cm, clip]{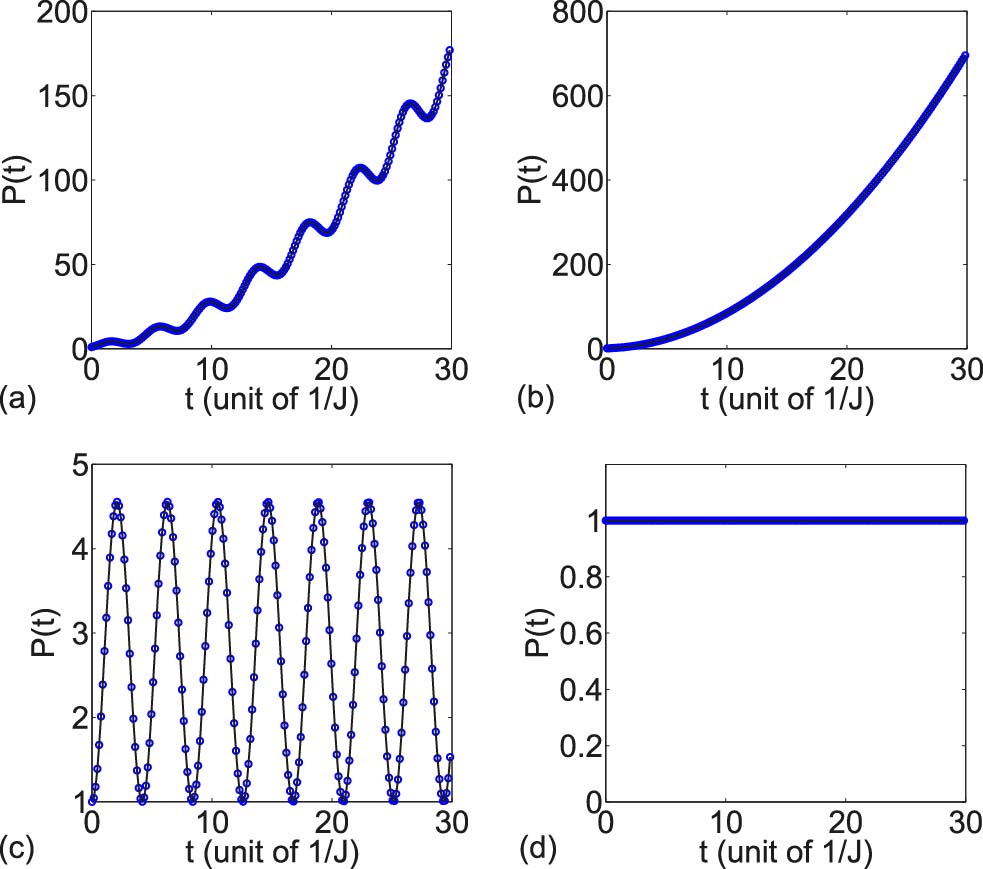}
\caption{(Color online) Time evolution probability of the trimer at EP2.
Initial state is (a) $\left\vert \Psi (0)\right\rangle =[1,0,0]^{T}$, (b) $\left\vert \Psi (0)\right\rangle =(1/\protect\sqrt{2})[1,-1,0]^{T}$, (c) $\left\vert \Psi (0)\right\rangle =(1/\protect\sqrt{3})[1,1,1]^{T}$, and (d) $\left\vert \Psi (0)\right\rangle =(1/\protect\sqrt{2})[1,0,1]^{T}$. The
increase in (a, b) is quadratic, and the period in (a, c) is $4\protect\pi /3J$. The trimer parameters are $\protect\kappa =1/2$, $J=1$, and $\protect\gamma =\protect\sqrt{3}/2$ at (a-c) $\Phi =-\protect\pi /3$ and at (d) $\Phi =\protect\pi /3$. The black lines (blue crosses) are the analytical
(numerical) results.} \label{fig6}
\end{figure}

The EP2 in the trimer ring occurs when the effective magnetic flux $\Phi
\neq 2n\pi +\pi /2$ ($n\in
\mathbb{Z}
$). The critical value for the balanced gain and loss is revealed in Eq. (%
\ref{EP2}), where we denote $E_{1}$ as the energy of the normal eigenstate,
and $E_{2}$ as the coalesced eigenenergy at EP2, and they satisfy $%
E_{1}=-2E_{2}$. Considering effective magnetic flux $\Phi =-\pi /3$ as an
example, the trimer ring is at the EP2 for the coupling strengths of $J=1$
and $\kappa =1/2$, and the critical balanced gain and loss of $\gamma _{%
\mathrm{c}}=\sqrt{3}/2$. Here, the three eigenenergies are one $E_{1}=-1$,
and two coalesced energy $E_{2}=1/2$. Generally, the probability of an
initial state increases quadratically under an oscillation [Fig.~\ref{fig6}%
(a)]. The oscillation is attributed to the contribution of two different
eigenstates (with different real energies $E_{1}$ and $E_{2}$), the increase
results from the two coalesced states (with eigenenergy $E_{2}$).
Characteristic dynamics emerge in some typical cases, which are determined
by the special initial states. When the initial state satisfies $\Psi
_{1}(0)+\Psi _{2}(0)+\Psi _{3}(0)=0$, the contribution of the normal
eigenstate $E_{1}$ is zero and the probability increases monotonously
without oscillation [Fig.~\ref{fig6}(b)]; By contrast, when the initial
state satisfies $\Psi _{1}(0)=\Psi _{2}(0)=\Psi _{3}(0)$, this relation
leads to the disappearance of the associated states' contribution. Thus, the
state probability oscillates in a range rather than increasing with time.
The exact expression is $P(t)=[25-16\cos \left( 3t/2\right) ]/9$, as
depicted in Fig.~\ref{fig6}(c), the period is determined by the energy $%
T=2\pi /E_{2}=4\pi /3$. Figure~\ref{fig6}(d) shows a probability
conservation, which occurs at $\Phi =\pi /3$ for $\Psi (0)=(1/\sqrt{2})[%
\begin{array}{ccc}
1 & 0 & 1%
\end{array}%
]^{T}$. In this situation, the probabilities at the gain and loss sites
remain the same in the time evolution process, and they oscillate at period $%
T=4\pi /3$, but the probability for all three sites conserves.

At exceptional points, the system has eigenstates and associated states. All
the eigenstates are also $\mathcal{PT}$-symmetric at exceptional points,
while the number of associated states is one less than the coalesced states.
In other words, the number of associated states is the same as the number of
eigenstates lacking at exceptional points. For the time evolution dynamics,
the power oscillation is typically found in the exact $\mathcal{PT}$%
-symmetric region; however, when the initial state is only eigenstates
relevant at the exceptional point (i.e., not related to the associated
state), the probability oscillates instead of increasing.

\section{Scattering properties}

\label{SS} In this section, we studied the scattering properties of the
trimers. We considered the trimer embedded in a uniform chain as the
scattering center. The coupling strength in the chain is $\kappa $, and the
left lead is $-\kappa \sum_{j=-\infty }^{0}(b_{j}^{\dagger
}b_{j+1}+b_{j+1}^{\dagger }b_{j})$, whereas the right lead is $-\kappa
\sum_{j=1}^{+\infty }(b_{j}^{\dagger }b_{j+1}+b_{j+1}^{\dagger }b_{j})$. The
operator $b_{j}^{\dagger }$ ($b_{j}$) is the creation (annihilation)
operator in the leads. The input lead is connected to the gain site, whereas
the output lead is connected to the lossy site, and the connection
Hamiltonian is $-\kappa (b_{0}^{\dagger }a_{1}+b_{1}^{\dagger }a_{3}+\mathrm{%
h.c.})$. Additionally, the scattering wave function in the input lead is
denoted as $\langle j\left\vert \psi _{k}\right\rangle =e^{ikj}+re^{-ikj}$,
whereas in the output lead it is denoted as $\langle j\left\vert \psi
_{k}\right\rangle =te^{ikj}$, where $j$ is the site number in the input and
output lead. $t_{L(R)}$ is the transmission coefficient, and $r_{L(R)}$ is
the reflection coefficient for the left (right) side input. The transmission
and reflection probabilities are denoted as $T_{L(R)}=|t_{L(R)}|^{2}$ and $%
R_{L(R)}=|r_{L(R)}|^{2}$.

We consider the trimer chain embedded in the lead as a scattering center.
The scattering coefficients for input wave vector $k$ are
\begin{eqnarray}
t_{L} &=&\frac{i\sin k}{i\sin k-e^{2ik}\gamma ^{2}\cos k}=t_{R}, \\
r_{L} &=&\frac{\gamma ^{2}\cos k+\gamma \sin \left( 2k\right) }{i\sin
k-e^{2ik}\gamma ^{2}\cos k}, \\
r_{R} &=&\frac{\gamma ^{2}\cos k-\gamma \sin \left( 2k\right) }{i\sin
k-e^{2ik}\gamma ^{2}\cos k}.
\end{eqnarray}

\begin{figure}[tb]
\includegraphics[bb=0 0 360 500, width=8.4 cm, clip]{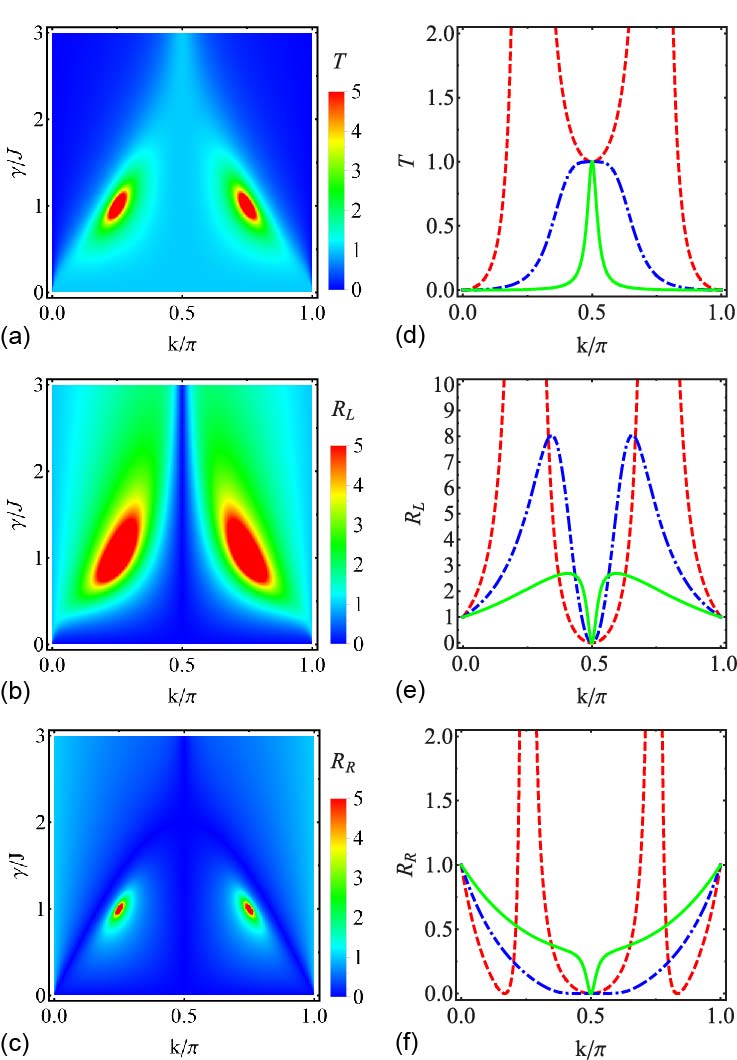}
\caption{(Color online) Transmission and reflection probability. (a,d) The
transmissions, (b,e) the left reflection, and (c,f) the right reflection.
(d,e,f) The red dashed, blue dash-dotted, and green solid curves represent $\protect\gamma=\protect\kappa$, $\protect\gamma=2\protect\kappa$, and $\protect\gamma=4\protect\kappa$, respectively. The divergence is at $\protect\gamma=\protect\kappa$ for $k=\protect\pi/4$ and $3\protect\pi/4$.}
\label{fig7}
\end{figure}

\begin{figure}[tb]
\includegraphics[bb=0 0 500 410, width=5 cm, clip]{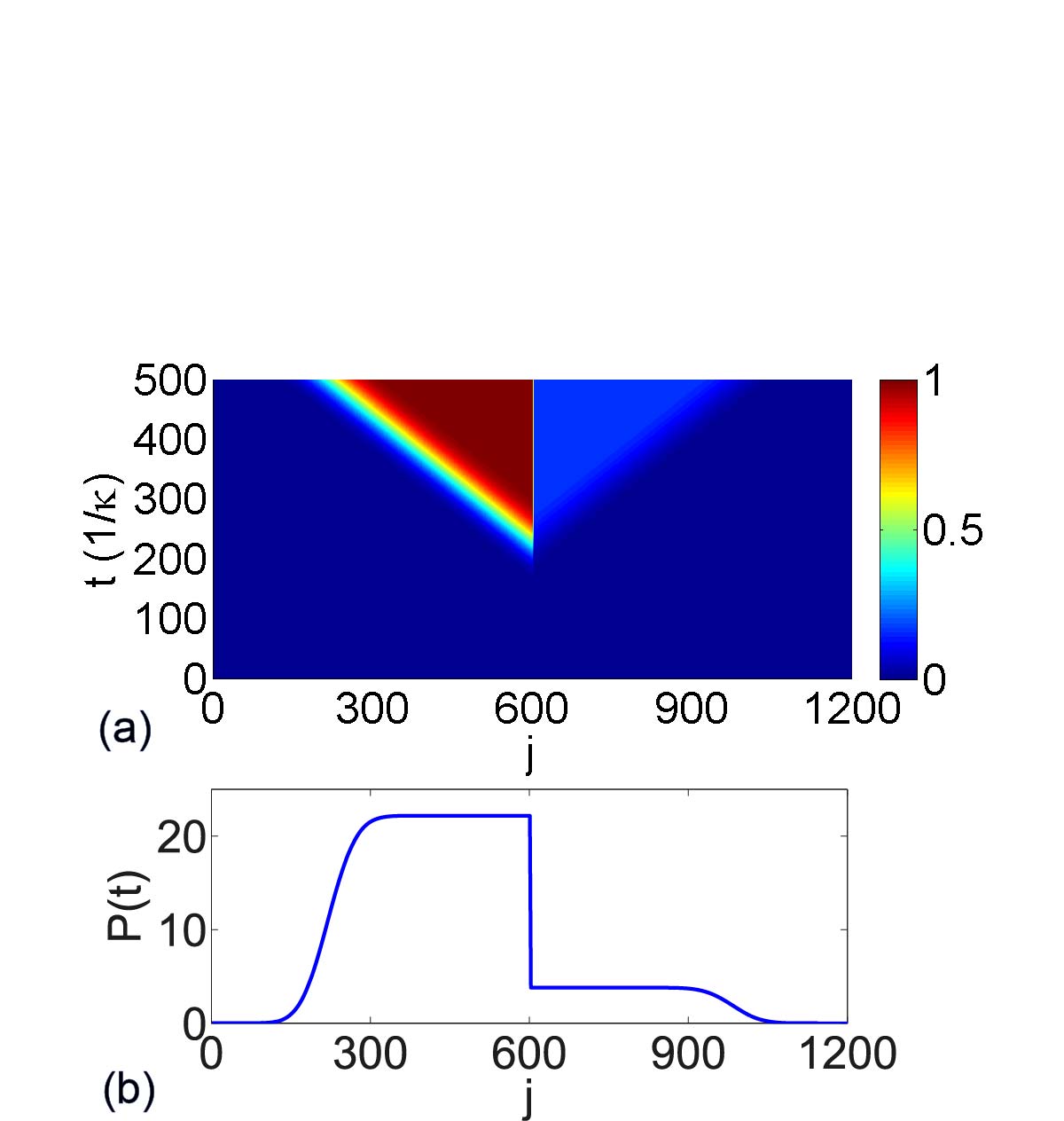}
\caption{(Color online) Wave emission dynamics at spectral singularity for
the trimer chain. A Gaussian wave packet with $\protect\alpha =0.02$
centered at $N_{\mathrm{c}}=900$ is traveling leftward initially, and the
trimer chain is at the spectral singularity for $\protect\gamma =\protect\kappa $, $\protect\kappa =\protect\pi /4$. The left-traveling wave height
is $\protect\sqrt{\protect\pi }/(4\protect\alpha )$.} \label{fig8}
\end{figure}

From the expression of the scattering coefficient, the transmission is
symmetric (i.e., $T=T_{L}=T_{R}$). The transmission coefficients are $%
t_{L}=t_{R}=1$, and the reflection coefficients are $r_{L}=r_{R}=0$ for
input wave vector $k=\pi /2$. Thus, the waves are completely unaffected by
the gain and loss. In Fig.~\ref{fig7}, the transmission and reflection
probabilities are plotted. The transmission is larger than the unity for a
wide region of input wave vectors, approximately in the weak gain and loss
region $\gamma /J<1$. For strong gain and loss ($\gamma /J>2$), the
transmission is less than the unity and close to zero in a wide region away
from the input wave vector $k=\pi /2$ [Fig.~\ref{fig7}(a) and Fig.~\ref{fig7}%
(d)]. Moreover, the left reflection is less than the unity when the wave
vector approaches $k=\pi /2$, or at small gain and loss region ($\gamma
/J\ll 1$); otherwise, the left transmission is substantially larger than the
unity [Fig.~\ref{fig7}(b) and Fig.~\ref{fig7}(e)]. The right reflection has
an opposite law to the left reflection; specifically, the right reflection
is usually less than the unity except near the spectral singularities~\cite%
{AliPRL,AliPRA13,LGR}. The right reflection also only changes dramatically
in close vicinity to the spectral singularities [Fig.~\ref{fig7}(c) and Fig.~%
\ref{fig7}(f)]. For large $\gamma /J\gg 1$, $R_{L}R_{T}\approx 1$ at $k$
deviated from $\pi /2$ and $R_{L}R_{T}\approx 0$ at $k$ near $\pi /2$, as
indicated by the green lines in Fig.~\ref{fig7}(e) and~\ref{fig7}(f). In the
vicinity of $k=\pi /2$, the transmission is close to the unity and the left
and right reflections are near zero.

The scattering system has one spectral singularity at
\begin{equation}
\gamma =\kappa ,
\end{equation}%
for input wave vectors $k=\pi /4,3\pi /4$ (we assume $\kappa ,\gamma >0$)
and the waves correspond to energy $E=-\sqrt{2}\kappa $ and velocity $\sqrt{2%
}\kappa $. At spectral singularity, all of the scattering coefficients
diverge, $t_{L,R}$, $r_{L,R}\rightarrow \infty $. The emission is
asymmetric, and the ratio for the right-traveling and left-traveling waves
is $3-2\sqrt{2}\approx 0.17$. As we numerically simulated the wave emission
process depicted in Fig.~\ref{fig8}. The initial state is a Gaussian wave
packet, $\left\vert \Psi \left( 0\right) \right\rangle =\left( \sqrt{\pi }%
/\alpha \right) ^{-1/2}\sum_{j}e^{-(\alpha ^{2}/2)\left( j-N_{\mathrm{c}%
}\right) ^{2}}e^{ikj}\left\vert j\right\rangle $ centered at $N_{\mathrm{c}}$%
. At the spectral singularities, wave emission toward both sides occurs
after the wave packet reaches the trimer scattering center, and it forms a
stepped platform. The platform heights for the transmitted part are $\sqrt{%
\pi }/(4\alpha )$, including the right-traveling wave for the left side
input and the left-traveling wave for the right side input. Moreover, after
being scattered by the trimer, the left-traveling wave height is $(3+2\sqrt{2%
})\sqrt{\pi }/(4\alpha )$, and the right-traveling wave height is $\sqrt{\pi
}/(4\alpha )$ (for left input); by contrast, the left-traveling wave height
is $\sqrt{\pi }/(4\alpha )$, and the right-traveling wave height is $(3-2%
\sqrt{2})\sqrt{\pi }/(4\alpha )$ (for the right input). These are revealed
by the contour plot in Fig.~\ref{fig8} (a), where the wave emission platform
with large height was renormalized to unity. The contour of a left input
centered at $N_{\mathrm{c}}=300$, which is the same as a contour of a right
input centered at $N_{\mathrm{c}}=900$. The initial Gaussian wave packet is
centered at $N_{\mathrm{c}}=900$ in the plot, and it moves from right to
left at a velocity $\sqrt{2}\kappa $. Fig.~\ref{fig8}(b) details the plotted
probability distribution of the wave function at $t=500/\kappa $, where the
asymmetric platform with different heights is clearly observed.

\begin{figure}[tb]
\includegraphics[bb=0 0 370 510, width=8.4 cm, clip]{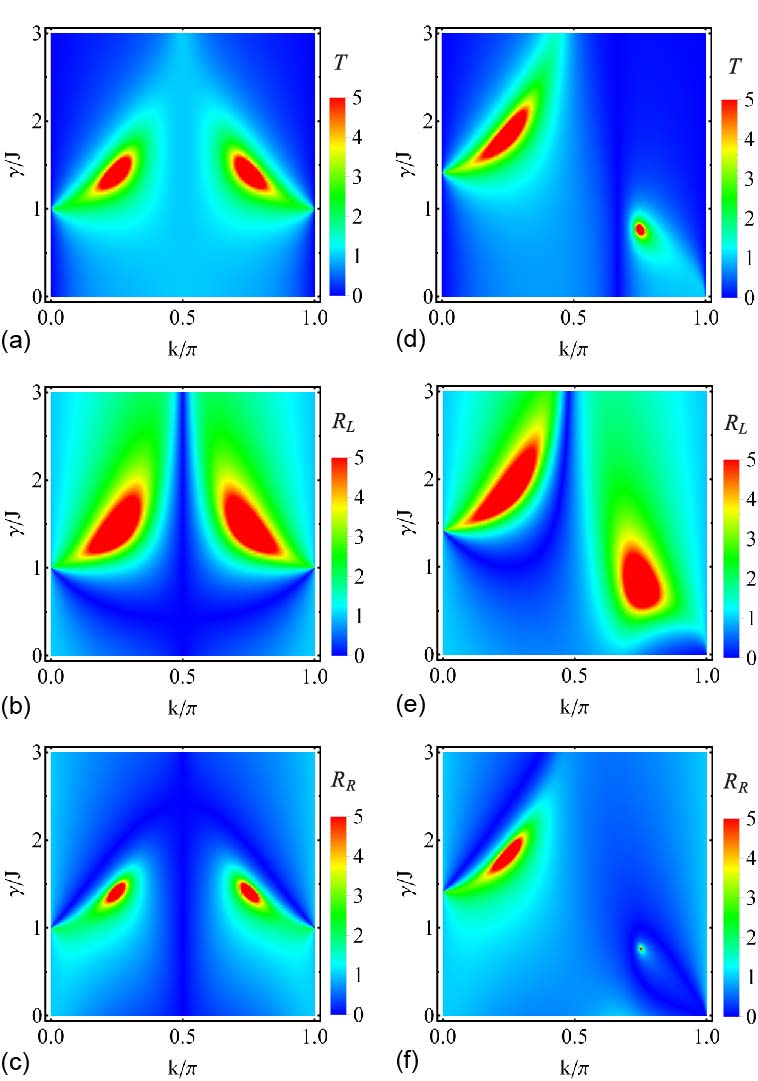}
\caption{(Color online) Transmission and reflection probability of the
uniform trimer ring ($J=\protect\kappa$) for (a-c) $\Phi=\protect\pi/2$ and
(d-f) $\Phi=0$.} \label{fig9}
\end{figure}

For the trimer ring embedded in the lead as a scattering center, the
scattering coefficients for the input wave vector $k$ are%
\begin{eqnarray}
t_{L} &=&\frac{i\sin k\left( e^{-i\Phi }+2\cos k\right) }{ie^{-2ik}\sin
k+\cos \Phi +\cos k\left( 1-\gamma ^{2}\right) }, \\
r_{L} &=&\frac{i\sin k\left( e^{-2ik}-2i\gamma \cos k\right) }{ie^{-2ik}\sin
k+\cos \Phi +\cos k\left( 1-\gamma ^{2}\right) }-1,
\end{eqnarray}%
\begin{eqnarray}
t_{R} &=&\frac{i\sin k\left( e^{i\Phi }+2\cos k\right) }{ie^{-2ik}\sin
k+\cos \Phi +\cos k\left( 1-\gamma ^{2}\right) }, \\
r_{R} &=&\frac{i\sin k\left( e^{-2ik}+2i\gamma \cos k\right) }{ie^{-2ik}\sin
k+\cos \Phi +\cos k\left( 1-\gamma ^{2}\right) }-1.
\end{eqnarray}%
where the coupling strengths satisfy $J=\kappa $. From the expressions of
the scattering coefficients, the transmission is symmetric (i.e., $%
T=T_{L}=T_{R}$). Figure~\ref{fig9} illustrates the plotted transmission and
reflection probabilities. In the region $k\in \lbrack 0,\pi ]$, the
scattering probability is symmetrical at $k=\pi /2$ for effective magnetic $%
\Phi =2n\pi +\pi /2$ ($n\in
\mathbb{Z}
$) [Fig.~\ref{fig9}(a-c)], where the trimer ring possesses chiral symmetry
and its spectrum is symmetrical at zero energy (i.e., the energy $-2\cos
(\pi /2)$). In other magnetic fluxes, such as $\Phi =0$, the chiral symmetry
of the trimer ring breaks, and the scattering probabilities are asymmetric
at $k=\pi /2$ [Fig.~\ref{fig9}(d-f)]. When $k=\pi /2$, the reflection and
transmission are identical, both symmetrical and $\gamma $-independent, $%
T_{L}=R_{L}=T_{R}=R_{R}=\left( \cos ^{2}\Phi +1\right) ^{-1}$. The
divergence of the scattering coefficients occurs only for wave vectors $%
k=\pi /4$ and $3\pi /4$. At $k=\pi /4$, the transmissions ($t_{L}$ and $%
t_{R} $) are larger than zero under any balanced trimer gain and loss $%
\gamma $ or when the effective magnetic flux $\Phi $ enclosed. The
scattering coefficients divergence emerges at
\begin{equation}
\gamma ^{2}=\sqrt{2}\cos \Phi +2,
\end{equation}%
and the spectral singularity emerges in the region $\sqrt{2-\sqrt{2}}%
\leqslant |\gamma /\kappa |\leqslant \sqrt{2+\sqrt{2}}$. Notably, the trimer
acts as a wave emitter at the spectral singularities. The wave emission is
asymmetric, and the contrast ratio of left-traveling wave emission over
right-traveling wave emission is $(\sqrt{2}\gamma -1)/(\sqrt{2}\gamma +1)$,
which is obtained by comparing the transmission and reflection
probabilities. The contrast ratio less than one is related to left side with
the gain resonator having a larger emission rate. For example, at $\gamma =%
\sqrt{2}$, $\Phi =\pi /2$, we have $T_{L}/R_{L}=1/3$, $T_{R}/R_{R}=3$. This
indicates that the lasing to different directions have different
intensities, similar to the trimer chain. The lasing generates three times
as many left-traveling waves than does the right-traveling wave. For either
the trimer chain or the trimer ring, the asymmetric lasing intensity are
both stronger right-traveling waves.

\section{Conclusion}

\label{Summary} Two $\mathcal{PT}$-symmetric trimer systems, comprising an
open trimer chain and a closed trimer ring were examined in this study.
First, we investigated the spectra of the trimer systems and obtained the $%
\mathcal{PT}$-symmetric phase diagram. we found that the nontrivial magnetic
flux suppresses the $\mathcal{PT}$ transition. For the trimer chain, the $%
\mathcal{PT}$-symmetric phase transition point is an EP3 with three states
coalescence; for the trimer ring, except for the EP3, the $\mathcal{PT}$%
-symmetric phase transition point can be an EP2 with two states coalescence
as long as an effective magnetic flux breaks the chiral symmetry of the
trimer ring. Second, we studied the time evolution at the exceptional point
in detail, and determined that the state probability can change in four ways
when the trimer ring is at EP2: (i) unchanged, (ii) oscillation, (iii) power
law increase, and (iv) power law increase with oscillation. These occur
because of the nonzero coalesced eigenstate and the third normal eigenstate.
Finally, we investigated the scattering properties of the trimer systems
with the gain and loss sites embedded in the lead, and calculated the
transmissions and reflections for the left- and right-side inputs. Notably,
the transmission is symmetric and the reflection is asymmetric. We also
found the spectral singularities, at which wave emission is asymmetric. The
critical dynamics at exceptional points and the scattering properties at
spectral singularities may be useful in quantum metrology in the future.

\textbf{Note added:} We become aware of a related work, the dynamics in the
exact and broken $\mathcal{PT}$-symmetric phases were investigated in an
open trimer chain~\cite{ZHWang}.

\acknowledgments We acknowledge the support of National Natural Science
Foundation of China (Grant No. 11605094) and the Tianjin Natural Science
Foundation (Grant No. 16JCYBJC40800).

\appendix

\section{Dynamics at exceptional points}

\label{Appendix}

For the trimer ring at EP3, the Hamiltonian at critical gain and loss $%
\gamma _{\mathrm{c}}=\sqrt{J^{2}+2\kappa ^{2}}$ is in the form of%
\begin{equation}
H=\left(
\begin{array}{ccc}
i\sqrt{J^{2}+2\kappa ^{2}} & -\kappa & iJ \\
-\kappa & 0 & -\kappa \\
-iJ & -\kappa & -i\sqrt{J^{2}+2\kappa ^{2}}%
\end{array}%
\right) .
\end{equation}%
The Hamiltonian can be tranformed to $H=VhV^{-1}$ with the transformation
\begin{equation}
V=\left(
\begin{array}{ccc}
-\kappa ^{2} & i\sqrt{J^{2}+2\kappa ^{2}} & 1 \\
-i\kappa \left( J+\sqrt{J^{2}+2\kappa ^{2}}\right) & -\kappa & 0 \\
\kappa ^{2} & iJ & 0%
\end{array}%
\right) ,
\end{equation}%
and $h$ is a $3\times 3$ Jordan block with diagonal elements $\lambda =0$,
\begin{equation}
h=\left(
\begin{array}{ccc}
\lambda & 1 & 0 \\
0 & \lambda & 1 \\
0 & 0 & \lambda%
\end{array}%
\right) .
\end{equation}

The Schr\"{o}dinger equations are
\begin{equation}
\frac{i\mathrm{d}}{\mathrm{d}t}\left(
\begin{array}{c}
\Psi _{1} \\
\Psi _{2} \\
\Psi _{3}%
\end{array}%
\right) =H\left(
\begin{array}{c}
\Psi _{1} \\
\Psi _{2} \\
\Psi _{3}%
\end{array}%
\right) ,
\end{equation}%
substituting $H=VhV^{-1}$, we obtain%
\begin{equation}
\frac{i\mathrm{d}}{\mathrm{d}t}V^{-1}\left(
\begin{array}{c}
\Psi _{1} \\
\Psi _{2} \\
\Psi _{3}%
\end{array}%
\right) =hV^{-1}\left(
\begin{array}{c}
\Psi _{1} \\
\Psi _{2} \\
\Psi _{3}%
\end{array}%
\right) ,
\end{equation}%
by setting $\psi =V^{-1}\Psi $, the Schr\"{o}dinger equations are reduced to
differential equations of $\psi $, in the form of%
\begin{equation}
\frac{i\mathrm{d}}{\mathrm{d}t}\left(
\begin{array}{c}
\psi _{1} \\
\psi _{2} \\
\psi _{3}%
\end{array}%
\right) =h\left(
\begin{array}{c}
\psi _{1} \\
\psi _{2} \\
\psi _{3}%
\end{array}%
\right) ,
\end{equation}%
which are%
\begin{eqnarray}
\frac{i\mathrm{d}\psi _{1}}{\mathrm{d}t} &=&\lambda \psi _{1}+\psi _{2}, \\
\frac{i\mathrm{d}\psi _{2}}{\mathrm{d}t} &=&\lambda \psi _{2}+\psi _{3}, \\
\frac{i\mathrm{d}\psi _{3}}{\mathrm{d}t} &=&\lambda \psi _{3}.
\end{eqnarray}%
From the last equation, we get%
\begin{equation}
\psi _{3}=c_{3}e^{-i\lambda t},
\end{equation}%
thus, we have $i\mathrm{d}\psi _{2}/\mathrm{d}t=\lambda \psi
_{2}+c_{3}e^{-i\lambda t}$, and then we get
\begin{equation}
\psi _{2}=c_{2}e^{-i\lambda t}+\left( -it\right) c_{3}e^{-i\lambda t},
\end{equation}%
consequently, we have $i\mathrm{d}\psi _{1}/\mathrm{d}t=\lambda \psi
_{1}+c_{2}e^{-i\lambda t}+\left( -it\right) c_{3}e^{-i\lambda t}$ and
\begin{equation}
\psi _{1}=c_{1}e^{-i\lambda t}+\left( -it\right) c_{2}e^{-i\lambda
t}-(t^{2}/2)c_{3}e^{-i\lambda t}.
\end{equation}

The obtained wave function $\psi \left( t\right) $ is%
\begin{equation}
\psi \left( t\right) =e^{-i\lambda t}\left(
\begin{array}{c}
c_{1}+\left( -it\right) c_{2}-(t^{2}/2)c_{3} \\
c_{2}+\left( -it\right) c_{3} \\
c_{3}%
\end{array}%
\right) .
\end{equation}%
Thus, the time evolution state is $\Psi \left( t\right) =V\psi \left(
t\right) $. The coefficients $c_{1,2,3}$ are determined from the initial
state. At $t=0$, we have $\Psi \left( 0\right) =V\psi \left( 0\right) $,
therefore, the initial state is expressed as%
\begin{equation}
\Psi \left( 0\right) =\left(
\begin{array}{c}
\Psi _{1} \\
\Psi _{2} \\
\Psi _{3}%
\end{array}%
\right) =V\left(
\begin{array}{c}
c_{1} \\
c_{2} \\
c_{3}%
\end{array}%
\right) ,
\end{equation}%
the coefficients satisfy $(%
\begin{array}{ccc}
c_{1} & c_{2} & c_{3}%
\end{array}%
)^{T}=V^{-1}(%
\begin{array}{ccc}
\Psi _{1} & \Psi _{2} & \Psi _{3}%
\end{array}%
)^{T}$, and the time evolution state $\Psi \left( t\right) $ is determined,
as%
\begin{equation}
\Psi \left( t\right) =e^{-i\lambda t}V\left(
\begin{array}{c}
c_{1}+\left( -it\right) c_{2}-(t^{2}/2)c_{3} \\
c_{2}+\left( -it\right) c_{3} \\
c_{3}%
\end{array}%
\right) .
\end{equation}%
Here, $\lambda =0$ for the three states coalescence in the trimer ring.

For the trimer ring at EP2 at magnetic flux $\Phi =\pi /3$, the coupling
strengths are $\kappa =1$ and $J=1/2$, at critical gain and loss $\gamma _{%
\mathrm{c}}=\sqrt{3}/2$, the Hamiltonian is in the form of%
\begin{equation}
H=\left(
\begin{array}{ccc}
i\sqrt{3}/2 & -1/2 & -e^{i\pi /3} \\
-1/2 & 0 & -1/2 \\
-e^{-i\pi /3} & -1/2 & -i\sqrt{3}/2%
\end{array}%
\right) ,
\end{equation}%
The Hamiltonian can be tranformed to $H=VhV^{-1}$ with%
\begin{equation}
V=\left(
\begin{array}{ccc}
1-\frac{2}{3}i\sqrt{3} & \frac{1}{2}i\sqrt{3} & 2+\frac{2}{3}i\sqrt{3} \\
1-\frac{2}{3}i\sqrt{3} & -i\sqrt{3} & -1+\frac{2}{3}i\sqrt{3} \\
1-\frac{2}{3}i\sqrt{3} & \frac{1}{2}i\sqrt{3} & -1+\frac{2}{3}i\sqrt{3}%
\end{array}%
\right) ,
\end{equation}%
and
\begin{equation}
h=\left(
\begin{array}{ccc}
-1 & 0 & 0 \\
0 & \frac{1}{2} & 1 \\
0 & 0 & \frac{1}{2}%
\end{array}%
\right) .  \label{h}
\end{equation}

At $\Phi =-\pi /3$, $\kappa =1$, $J=1/2$, and $\gamma _{\mathrm{c}}=\sqrt{3}%
/2$, the trimer ring is still an EP2, a transformation is
\begin{equation}
V=\left(
\begin{array}{ccc}
1-\frac{2}{3}i\sqrt{3} & \frac{1}{2}i\sqrt{3} & 2+\frac{2}{3}i\sqrt{3} \\
1 & 0 & -1 \\
1+\frac{2}{3}i\sqrt{3} & -\frac{1}{2}i\sqrt{3} & -1-\frac{2}{3}i\sqrt{3}%
\end{array}%
\right) ,
\end{equation}%
and $h$ is identical with Eq. (\ref{h}). The time evolution is calculated by
the same method introduced in this Appendix.

\end{document}